\documentclass[lettersize,journal]{IEEEtran}

\usepackage{amssymb}
\usepackage{amsmath}
\usepackage{graphicx} 
\usepackage{cite}
\usepackage{txfonts}
\usepackage{mathrsfs} 
\usepackage{bbm}
\usepackage{fontenc}
\usepackage{siunitx}
\usepackage[caption=false,font=footnotesize]{subfig}
\usepackage{tabularx}
\usepackage{multirow}
\usepackage{diagbox}

\newtheorem{proposition}{Proposition}
\newtheorem{remark}{Remark}
\newtheorem{lemma}{Lemma}
\usepackage[ruled,vlined]{algorithm2e}
\usepackage{arydshln}
\def \tr {\mathrm{tr}}
\usepackage[normalem]{ulem}

\begin{document}
\title{Nonlinear Power Amplifier-Resilient Cell-Free Massive MIMO: A Joint Optimization Approach}
%\title{Joint User Association and Power Control for Nonlinear PA-Resilient Cell-Free Massive MIMO}

\author{Wei~Jiang,~\IEEEmembership{Senior~Member,~IEEE,}
        and Hans~Dieter~Schotten,~\IEEEmembership{Member,~IEEE}% <-this % stops a space
\thanks{Manuscript received xxx xxx, 2025; accepted December 7th, 2025. The associated editor to coordinate the review of this letters is Prof. Min-Kung Chang. (\textit{Corresponding author: Wei Jiang (e-mail: wei.jiang@dfki.de)})}
\thanks{W. Jiang and H. D. Schotten are with German Research Center for Artificial Intelligence (DFKI), Kaiserslautern, Germany, and are also with the University of Kaiserslautern (RPTU), Germany. This work was supported by the German Federal Ministry of Education and Research (BMBF) through \emph{Open6G-Hub} (Grant no.  \emph{16KISK003K}) and \emph{Open6GHub+} (Grant no.  \emph{16KIS2402K}) projects.}
}

\markboth{IEEE Wireless Communications Letters,~Vol.~x, No.~x, YY~2026}%
{Jiang \MakeLowercase{\textit{et al.}}: Nonlinear Power Amplifier-Resilient Cell-Free Massive MIMO}
\maketitle

\begin{abstract}
   This letter {proposes an analytical model to study} analyzes the effects of power amplifiers (PAs) on the downlink of cell-free massive MIMO systems. We model signal transmission incorporating nonlinear PA distortion and derive a unified spectral efficiency (SE) expression applicable to arbitrary precoding schemes. To combat PA-induced performance {loss}, a {tractable} joint optimization approach, {as well as its low-complexity alternative,} for {user-centric clustering} and max-min power control is proposed {based on a conservative approximation. } 
\end{abstract}
\begin{IEEEkeywords}
Cell-free massive MIMO,   max-min power control, nonlinear power amplifier, user association, user fairness
\end{IEEEkeywords}

\IEEEpeerreviewmaketitle

\section{Introduction}

\IEEEPARstart{C}{ell-free} (CF) massive multi-input multi-output (CF-mMIMO) has garnered significant attention recently.  Nevertheless, power amplifiers (PAs)—critical components in wireless transmitters—introduce nonlinear signal distortion, particularly when operating near their saturation point. % This nonlinearity distorts the transmitted signal, affecting the desired signal strength and introducing additional interference. 
Ignoring this nonlinearity can lead to overestimated system performance and cause bias into algorithm design.

To date, only a few existing works provide initial insights. Mokhtari et al. \cite{Ref_mokhtari2021pad} quantified sum-rate degradation under PA distortion using maximum ratio (MR) precoding with equal power allocation. Subsequent studies by Jadidi et al. \cite{Ref_jadidi2024uplinkPA} and Khoueini et al. \cite{Ref_khoueini2025downlinkPA} investigated uplink and downlink impacts, respectively, assuming single-antenna access points (APs) and MR precoding. While these studies offer valuable theoretical insights, several key dimensions remain unexplored, e.g.,
\begin{itemize}
    \item \textit{Scalability limits}: \cite{Ref_mokhtari2021pad, Ref_jadidi2024uplinkPA, Ref_khoueini2025downlinkPA} use conventional CF configuration where all APs serve all users, neglecting the scalability gain of user-centric dynamic clustering \cite{Ref_buzzi2020usercentric};
    \item \textit{Precoding generality constraint}: \cite{Ref_mokhtari2021pad, Ref_jadidi2024uplinkPA, Ref_khoueini2025downlinkPA} focus solely on MR precoding, omitting state-of-the-art techniques such as zero-forcing (ZF) and minimum mean square error (MMSE) precoding \cite{Ref_bjornson2020making}. 
\end{itemize}

To bridge these gaps, this letter presents a unified analytical framework and a novel design for PA-resilient CF-mMIMO systems. Specifically, we develop a downlink signal model that captures nonlinear PA distortion, and derive a spectral efficiency (SE) expression applicable to arbitrary linear precoding schemes. Our analysis reveals that the achievable SE under PA-induced distortion critically depends on both {user-centric clustering} and power control. {Leveraging this insight,} we propose a {tractable approximation} approach,  that {adaptively handles} {AP-user} association and {max–min fairness} transmission power to effectively suppress {the performance loss due to } nonlinear {PA} distortion. {The novelty of our work lies in moving beyond prior studies that focus either on PA-agnostic clustering \cite{Ref_buzzi2020usercentric, Ref_bjornson2020scalable} and power optimization \cite{ Ref_bjornson2020making, Ref_ngo2017cellfree} or on PA distortion analysis in non-clustering systems \cite{Ref_mokhtari2021pad, Ref_jadidi2024uplinkPA, Ref_khoueini2025downlinkPA }.}

The remainder of this letter is structured as follows. Section II introduces the system model. Section III presents the SE analysis under PA distortion. Section IV formulates the joint optimization design. Section V analyzes computational complexity and offers a low-complexity algorithm. %Finally, Section VI validates the gains via simulations.

\section{System Model}

We consider a CF-mMIMO system comprising $L$ APs, each equipped with $N_a$ co-located antennas, distributed throughout a coverage area to serve $K$ users, where the total number of antennas $M = L \times N_a\gg K$. Each user equipment (UE) is generally equipped with a single antenna. The sets of indices for APs and users are denoted by $\mathbb{L}= \{1,\ldots,L\}$ and $\mathbb{K}=\{1,\ldots,K\}$, respectively. 
The wireless channel from AP $l$, $\forall l\in \mathbb{L}$ to UE $k$, $\forall k \in \mathbb{K}$ is denoted by $\mathbf{h}_{kl}\in \mathbb{C}^{N_a}$. Adopting the block fading assumption, each \textit{coherent block}—a time-frequency span of $\tau_c$ symbols—maintains a quasi-static channel response. In practical deployments, closely spaced antennas at an AP exhibit spatial correlation due to their physical proximity. Hence, each coherence block applies an independent realization from \textit{correlated} Rayleigh fading, defined as $\mathbf{h}_{kl} \sim \mathcal{CN}(\mathbf{0}, \mathbf{R}_{kl} )$, where $\mathbf{R}_{kl}=\mathbb{E}[  \mathbf{h}_{kl} \mathbf{h}_{kl}^H ]$ 
stands for the spatial correlation matrix. 
Using linear  MMSE estimation \cite{Ref_bjornson2020making}, the estimated channel $\hat{\mathbf{h}}_{kl}$ follows a complex Gaussian distribution $\hat{\mathbf{h}}_{kl} \sim \mathcal{CN} \left(\mathbf{0}, p_u \tau_p \mathbf{R}_{kl} \boldsymbol{\Gamma}_{kl}^{-1}\mathbf{R}_{kl} \right)$,
where \(
     \boldsymbol{\Gamma}_{kl}=p_u\tau_p\sum \nolimits_{k'\in \mathcal{P}_k } \mathbf{R}_{k'l} + \sigma_z^2\mathbf{I}_{N_a}
     \), 
and $p_u$ denotes the maximal transmit power of the UEs, $\tau_p$ is the pilot sequence length, $\mathcal{P}_k$ is the set of users sharing the same pilot as user $k$, and $\sigma_z^2$ indicates the variance of thermal noise.
The estimation error, defined as $\tilde{\mathbf{h}}_{kl}=\mathbf{h}_{kl} - \hat{\mathbf{h}}_{kl}$, is attributed to both noise and pilot contamination, following $\mathcal{CN}(\mathbf{0}, \boldsymbol{\Theta }_{kl} )$, where the associated error covariance matrix is given by \(
\boldsymbol{\Theta}_{kl}=\mathbb{E}\left[ \tilde{ \mathbf{h}}_{kl} \tilde{\mathbf{h}}_{kl}^H \right]=\mathbf{R}_{kl} - p_u \tau_p \mathbf{R}_{kl} \boldsymbol{\Gamma}_{kl}^{-1}\mathbf{R}_{kl}
\).

\section{Downlink Transmission}

In the downlink, the data symbols intended for the $K$ users are assumed to be independent, zero-mean random variables with unit variance. These symbols are jointly expressed as $\mathbf{x} = [x_1,\ldots,x_K]^T$, satisfying $\mathbb{E}[\mathbf{x}\mathbf{x}^H] = \mathbf{I}_K$. Under the user-centric dynamic clustering strategy, each AP maintains a set of associated users, denoted by $\mathbb{K}_l$ for AP $l$, where {$\mathbb{K}_l \subseteq \mathbb{K} $. } Each AP spatially multiplexes the symbols of its associated users \( k \in \mathbb{K}_l \), producing
\begin{equation} \label{eQn:compositeTxSig_MR}
    \mathbf{s}_{l} = \sqrt{p_a}\sum\nolimits_{k\in \mathbb{K}} u_{kl}\sqrt{\eta_{kl}}\mathbf{w}_{kl} x_k,
\end{equation}
where $p_a$ denotes the AP power constraint,  $\eta_{kl}$ is the power coefficient for AP $l$ to user $k$, \( u_{kl} \) is the association indicator {for user-centric clustering}, which equals $1$ if user \( k \) is served by AP \( l \) (i.e., \( k \in \mathbb{K}_l \)), and $0$ otherwise. The vector $\mathbf{w}_{kl} \in \mathbb{C}^{N_a}$ denotes the precoding vector employed by AP $l$ for user $k$, with the normalization constraint $\mathbb{E}\left[\left|\mathbf{w}_{kl}\right|^2\right] = 1$.

\subsection{Nonlinear Power Amplification}
 A common approach to model PA nonlinearity is {the Bussgang decomposition \cite{Demir2020}}.
 %, which represents the PA output for a zero-mean Gaussian input signal as a linear component plus an uncorrelated distortion term. 
 For the feeding signal \(\mathbf{s}_l\), the amplified transmit signal  at AP \(l\)  is given by
\begin{equation}
    \mathbf{x}_{l} = \mathcal{G}(\mathbf{s}_l) = \alpha_l \sqrt{p_a}\sum\nolimits_{k\in \mathbb{K}} u_{kl}\sqrt{\eta_{kl}}\mathbf{w}_{kl} x_k + \mathbf{d}_l,
\end{equation}
where: \begin{itemize}
    \item $\mathcal{G}(\cdot)$: {A general nonlinear PA function.} 
    \item $\alpha_l$ is a complex linear gain, and $\alpha_l = \mathbb{E}\left[\mathbf{s}_l^H \mathcal{G}(\mathbf{s}_l)\right] / \mathbb{E}[\|\mathbf{s}_l\|^2]$.    
    \item $\mathbf{d}_l \sim \mathcal{CN}(0, \sigma_{d}^2\mathbf{I}_{N_a})$ is complex Gaussian distortion, uncorrelated with \(\mathbf{s}_l\) (i.e., \(\mathbb{E}[\mathbf{s}_l^H \mathbf{d}_l] = \mathbf{0}\)), where 
    \begin{equation} \label{GS_nonlinearpower}
        \sigma_{d}^2 = \beta_l \mathbb{E}\left[\|\mathbf{s}_l\|^2\right] = \beta_l p_a \sum\nolimits_{k \in \mathbb{K}} u_{kl}\eta_{kl}
    \end{equation}
    stands for distortion variance, and $\beta_l$ represents distortion-to-signal ratio.
\end{itemize}

Thus, the actual transmitted power is expressed by
\begin{align}
\mathbb{E}[\|\mathbf{x}_l\|^2] 
&= \mathbb{E}\left[\left\| \alpha_l \sqrt{p_a} \sum\nolimits_{k \in \mathbb{K}} u_{kl} \sqrt{\eta_{kl}} \mathbf{w}_{kl} x_k + \mathbf{d}_l \right\|^2\right] \notag \\
&= |\alpha_l|^2 p_a \sum\nolimits_{k \in \mathbb{K}} u_{kl} \eta_{kl} \mathbb{E}[\|\mathbf{w}_{kl}\|^2] + \mathbb{E}[\|\mathbf{d}_l\|^2], \label{eq:TxPower}
\end{align}
applying {the property of binary variables} $u_{kl}^2=u_{kl}$. 
The distortion power is  
\(
\mathbb{E}[\|\mathbf{d}_l\|^2] = \operatorname{tr}\left(\mathbb{E}[\mathbf{d}_l \mathbf{d}_l^H] \right) = \operatorname{tr}\left(\sigma_d^2 \mathbf{I}_{N_a} \right) = \beta_l p_a N_a \sum\nolimits_{k \in \mathbb{K}} u_{kl} \eta_{kl}
\), since \( \mathbb{E}[\|\mathbf{w}_{kl}\|^2] = 1 \).
Now, \eqref{eq:TxPower} simplifies to  
\(
    \mathbb{E}[\|\mathbf{x}_l\|^2] = \left(|\alpha_l|^2 p_a + \beta_l p_a N_a\right) \sum\nolimits_{k \in \mathbb{K}} u_{kl} \eta_{kl} \label{eq:TxPowerSimplified}
\).
To ensure the AP respects its power budget, the transmitted power must satisfy $\mathbb{E}[\|\mathbf{x}_l\|^2] \le p_a \label{eq:PowerConstraint}$, imposing the constraint:  
\begin{equation}
    \sum\nolimits_{k \in \mathbb{K}} u_{kl} \eta_{kl}\le \frac{1}{ |\alpha_l|^2  + \beta_l  N_a }, \quad\forall l \in \mathbb{L}.
\end{equation}

\subsection{Precoding Schemes} \label{sec:precoders}
To ensure scalability, each AP \( l \) independently performs local signal processing using its own channel estimates, denoted by \( \hat{\mathbf{H}}_l=\left[\hat{\mathbf{h}}_{1l},\cdots, \hat{\mathbf{h}}_{Kl}\right] \in \mathbb{C}^{N_a \times K} \). The applicable precoding schemes identified include:
\begin{enumerate}
    \item \textbf{MR}: This simple approach, a.k.a. conjugate beamforming, sets the precoding matrix at AP \( l \) as $\mathbf{V}_{l}^{mr}= \hat{\mathbf{H}}_{l}^*$ to maximize the desired signal strength.
    \item \textbf{ZF}: Aiming to cancel multi-user interference, the ZF precoder is the pseudo-inverse of \( \hat{\mathbf{H}}_l \), given by  \( \mathbf{V}_l^{\text{zf}} = \hat{\mathbf{H}}_l^* (\hat{\mathbf{H}}_l^T \hat{\mathbf{H}}_l^*)^{-1} \) so as to meet  \( \hat{\mathbf{H}}_l^T \mathbf{V}_l^{\text{zf}} = \mathbf{I}_K \).   This condition holds exclusively when the channel matrix satisfies the full-rank assumption, when \( \hat{\mathbf{H}}_l \) has full column rank, requiring \( N_a \geq K \).
    \item \textbf{Regularized ZF (RZF)}: In practice, due to scalability requirement, APs may be equipped with fewer antennas than users (\( N_a < K \)), rendering ZF inapplicable. To handle rank-deficient cases, regularized ZF is used, i.e., 
    \( 
    \mathbf{V}_l^{\text{rzf}} = \hat{\mathbf{H}}_l^* \left( \hat{\mathbf{H}}_l^T \hat{\mathbf{H}}_l^* + \sigma_z^2 \mathbf{I}_K \right)^{-1}
    \),
    where the regularization term \( \sigma_z^2 \mathbf{I}_K \) ensures matrix invertibility.
    \item \textbf{MMSE}: The balance between interference suppression and noise enhancement is achieved by minimizing the mean squared error between the transmitted symbol and the received signal. Its precoding matrix is given by $
        \mathbf{V}_l^{\text{mmse}} = \left( p_u \hat{\mathbf{H}}_l \mathbf{E}_l \hat{\mathbf{H}}_l^H + p_u \sum\nolimits_{k\in \mathbb{K} } \eta_{kl} \boldsymbol{\Theta}_{kl}  + \sigma_z^2 \mathbf{I}_{N_a} \right)^{-1} \hat{\mathbf{H}}_l$,
    where \( \mathbf{E}_l = \mathrm{diag}(\eta_{1l}, \dots, \eta_{Kl}) \) is a diagonal matrix of power coefficients.
\end{enumerate}

Let \( \mathbf{v}_{kl} \in \mathbb{C}^{N_a} \) denote the \( k \)-th column of \( \mathbf{V}_l \in \mathbb{C}^{N_a \times K} \), i.e., $\mathbf{v}_{kl} = \mathbf{V}_l(:,\,k)$.
The required precoding vector \(\mathbf{w}_{kl}\) is obtained by normalizing \(\mathbf{v}_{kl}\) as  
$\mathbf{w}_{kl} = \frac{\mathbf{v}_{kl}}{\|\mathbf{v}_{kl}\|},\quad \forall l \in \mathbb{L}, \forall k \in \mathbb{K}$, ensuring {it satisfies the normalization condition} \(\mathbb{E}\left[\|\mathbf{w}_{kl}\|^2\right] = 1\).

 \subsection{Performance Analysis}
The received signal at user \( k \) is  $y_k  =   \sum\nolimits_{l\in \mathbb{L} } \mathbf{h}_{kl}^T\mathbf{x}_l  +n_k  $, where noise \( n_k \sim \mathcal{CN}(0, \sigma_z^2) \). The expression can be expanded as
\begin{align} \label{eQn_downlinkModel} 
    y_k &= \sqrt{p_a} \sum_{l \in \mathbb{L}} \alpha_l \sum_{k' \in \mathbb{K}} u_{k'l} \sqrt{\eta_{k'l}} \mathbf{h}_{kl}^T \mathbf{w}_{k'l} x_{k'} + \sum_{l \in \mathbb{L}} \mathbf{h}_{kl}^T \mathbf{d}_l + n_k.
\end{align}

{Massive MIMO systems typically operate in time-division duplexing to avoid prohibitive downlink training overhead that scales with $M$. Therefore,} users {typically do} not know channel estimates due to the absence of downlink pilots, making coherent detection impractical. Instead, signal detection is performed using large-scale fading decoding (LSFD), where {the statistical mean \( \mathbb{E} [ \mathbf{w}_{kl}^H \mathbf{h}_{kl} ] \)}\footnote{{ Since the precoder $\mathbf{w}_{kl}$ is a function of the channel estimate $\hat{\mathbf{h}}_{kl}$ and is therefore uncorrelated with the channel estimation error $\tilde{\mathbf{h}}_{kl}$, it follows that $\mathbb{E} [ \mathbf{w}_{kl}^H \mathbf{h}_{kl} ]=\mathbb{E} [ \mathbf{w}_{kl}^H (\hat{\mathbf{h}}_{kl}+\tilde{\mathbf{h}}_{kl}) ]=\mathbb{E} [ \mathbf{w}_{kl}^H \hat{\mathbf{h}}_{kl} ]$. This expectation can be derived from channel statistics, see, e.g., \cite[Corol. 3]{Ref_bjornson2020scalable}.} }, $\forall k,l$, serves as an approximation. This statistical mismatch leads to an additional performance degradation known as channel uncertainty error.
To facilitate the derivation, \eqref{eQn_downlinkModel} is decomposed into 
\begin{align} \nonumber \label{eQn_DLGeneralSig}
    y_k = &  \underbrace{   \sqrt{p_a}\sum\nolimits_{l\in \mathbb{L}} \alpha_l u_{kl} \sqrt{\eta_{kl}}\mathbb{E} [\mathbf{h}_{kl}^T\mathbf{w}_{kl}]  x_k}_{\mathcal{S}_1:\:\text{desired signal over channel statistics}} \\ + & \underbrace{ 
    \sqrt{p_a}\sum\nolimits_{l\in \mathbb{L}} \alpha_l u_{kl} \sqrt{\eta_{kl}} \left(\mathbf{h}_{kl}^T\mathbf{w}_{kl}- \mathbb{E} [\mathbf{h}_{kl}^T\mathbf{w}_{kl}]\right) x_k  }_{\mathcal{J}_1:\:channel\:uncertainty\:error} \\  + & \nonumber
    \underbrace{  \sqrt{p_a}\sum\limits_{l\in \mathbb{L}} \alpha_l \sum\limits_{k'\in \mathbb{K} \backslash \{k\}} u_{k'l}\sqrt{\eta_{k'l}} \mathbf{h}_{kl}^T  \mathbf{w}_{{k'l}} x_{k'}
    }_{\mathcal{J}_2:\:inter-user\:interference} + \underbrace{\sum\nolimits_{l \in \mathbb{L}} \mathbf{h}_{kl}^T \mathbf{d}_l}_{\mathcal{J}_3: \text{PA distortion}} + n_k.
\end{align}
\begin{proposition}
    The achievable SE of user $k$ is $R_k=\mathbb{E} \Bigl[\log_2(1+\gamma_k) \Bigr]$, where {the expectation is over channel realizations} and the instantaneous effective SINR is 
\begin{equation} \label{eQn:DLSINR_2}
    \gamma_k = \frac{  \left|  \sum\nolimits_{l\in \mathbb{L}} \alpha_l u_{kl} \sqrt{\eta_{kl}}\mathbb{E} [\mathbf{h}_{kl}^T\mathbf{w}_{kl}] \right|^2  }{ \left\{ \begin{aligned}            
    & {\sigma_z^2}/{p_a} + \sum\nolimits_{k'\in \mathbb{K} } \left(      \sum\nolimits_{l\in\mathbb{L}} u_{k'l}  \eta_{k'l} |\alpha_l|^2\mathbb{E}[|  \mathbf{h}_{kl}^T  \mathbf{w}_{{k'l}}   |^2] \right)\\
     & {-}  \sum\limits_{l \in \mathbb{L}} u_{kl} \eta_{kl}|\alpha_l|^2  \left|\mathbb{E}[ \mathbf{h}_{kl}^T \mathbf{w}_{kl} ] \right|^2 {+} \sum_{l \in \mathbb{L}} \beta_l  \left( \sum_{k' \in \mathbb{K}} u_{k'l} \eta_{k'l} \right) \tr (\mathbf{R}_{kl})  \end{aligned} \right\} } 
\end{equation}
\end{proposition}
\begin{IEEEproof}
{Using the standard capacity lower bounds (cf. \cite[Prop. 3]{Ref_bjornson2020scalable}), the} SINR is given by $ \gamma_k  = \frac{|\mathcal{S}_1|^2}{\mathbb{E}\left[|\mathcal{J}_1+\mathcal{J}_2+\mathcal{J}_3|^2\right] + \sigma_z^2}$,
where the power gain of the desired signal is
\begin{equation}
    |\mathcal{S}_1|^2  = p_a \left| \sum\nolimits_{l\in \mathbb{L}} \alpha_l u_{kl} \sqrt{\eta_{kl}}\mathbb{E} [\mathbf{h}_{kl}^T\mathbf{w}_{kl}] \right|^2.
\end{equation}
Due to the independence among data symbols and signal distortion, $\mathcal{J}_1$, $\mathcal{J}_2$, and $\mathcal{J}_3$ in \eqref{eQn_DLGeneralSig} are uncorrelated. This implies that $\mathbb{E}\left[|\mathcal{J}_1 + \mathcal{J}_2+\mathcal{J}_3|^2\right]=\mathbb{E}\left[|\mathcal{J}_1|^2\right]+\mathbb{E}\left[|\mathcal{J}_2|^2\right]+\mathbb{E}\left[|\mathcal{J}_3|^2\right]$. We now compute the variance of $\mathcal{J}_1$ {(cf. \cite[Th. 1]{Ref_ngo2017cellfree})} as 
\begin{align}  \label{APPEQ1}
    \mathbb{E}\left[|\mathcal{J}_1|^2\right]  = &   p_a \sum\nolimits_{l \in \mathbb{L}}  u_{kl} \eta_{kl} |\alpha_l|^2 \mathbb{E}\left[ \left|  \mathbf{h}_{kl}^T\mathbf{w}_{kl}- \mathbb{E} [\mathbf{h}_{kl}^T\mathbf{w}_{kl}]  \right|^2 \right] \\ \nonumber
    = &   p_a \sum\nolimits_{l \in \mathbb{L}} u_{kl} \eta_{kl} |\alpha_l|^2 \left (  \mathbb{E}\left[ \left| \mathbf{h}_{kl}^T \mathbf{w}_{kl}\right|^2 \right] - \left|\mathbb{E}[ \mathbf{h}_{kl}^T \mathbf{w}_{kl} ] \right|^2 \right).
\end{align}
Next, the variance of $\mathcal{J}_2$ is computed as
\begin{align}  
    \mathbb{E}\left[|\mathcal{J}_2|^2\right]=  & p_a \sum\nolimits_{l\in\mathbb{L}} |\alpha_l|^2 \sum\nolimits_{k'\in \mathbb{K} \backslash \{k\} }   u_{k'l} \eta_{k'l} \mathbb{E}[|  \mathbf{h}_{kl}^T  \mathbf{w}_{{k'l}}   |^2]. 
\end{align}
Finally, \begin{align}  
    \mathbb{E}\left[|\mathcal{J}_3|^2\right]=   \sum_{l \in \mathbb{L}} \sigma_{d}^2 \mathbb{E}[\|\mathbf{h}_{kl}\|^2]  = p_a \sum_{l \in \mathbb{L}} \beta_l  \left( \sum_{k' \in \mathbb{K}} u_{k'l} \eta_{k'l} \right) \tr (\mathbf{R}_{kl}) . 
\end{align}
Applying $\mathcal{S}_1$, $\mathcal{J}_1$, $\mathcal{J}_2$, and $\mathcal{J}_3$ yields \eqref{eQn:DLSINR_2}.
\end{IEEEproof}

\section{Joint Optimization of User Association and Max-Min Power Allocation}
Analyzing  {\eqref{GS_nonlinearpower} and \eqref{eQn:DLSINR_2} reveal that user-centric AP association ($u_{kl}$) and power control ($\eta_{kl}$) decide the  variance of PA distortion and the resultant SINR.} To perserve uniformly high-quality service under PA-induced distortion, we propose jointly optimizing user association and max-min power allocation. The problem formulation is
\begin{align}   \label{eq:maxmin_problem}
    \max_{ \mathbf{U}, \; \boldsymbol{\eta}} \quad & \min_{k \in \mathbb{K}} \quad  \gamma_k\\ \nonumber
    \text{s.t.} \quad  & \left\{
    \begin{aligned}
    & \sum\nolimits_{k \in \mathbb{K}} u_{kl} \eta_{kl}\le \frac{1}{ |\alpha_l|^2  + \beta_l  N_a }, \ \forall l \in \mathbb{L}, \\
    & \eta_{kl} \geq 0, \quad u_{kl}\in \{0, 1\}, \quad \forall k \in \mathbb{K}, l \in \mathbb{L},
    \end{aligned}
    \right.
\end{align} where $\mathbf{U} = [u_{kl}]$ is the binary association matrix, and $\boldsymbol{\eta} = [\eta_{kl}]$ is the power allocation matrix. 
The numerator of $\gamma_k$ is a square of a sum of square roots, which is not convex in $\boldsymbol{\eta}$. To convexify this term, we introduce auxiliary variables $\nu_{kl}$, collectively denoted as $\boldsymbol{\nu} = \{\nu_{kl}\}_{k\in\mathbb{K},l\in\mathbb{L}}$, and impose the following constraint:
\begin{align}
    \label{eq:nu_eta_constraint}
    \nu_{kl}^2 \leq \eta_{kl}, \quad \forall k \in \mathbb{K}, l \in \mathbb{L}.
\end{align} 
{It can be equivalently expressed as a convex rotated second-order cone (SOC) constraint, given by}
\begin{align}
    \label{eq:rotated_soc_nu}
    \left\| \left[ 2 \nu_{kl};\: 1 - \eta_{kl} \right] \right\| \leq 1 + \eta_{kl}, \quad \forall k \in \mathbb{K}, l \in \mathbb{L}.
\end{align}

{This standard reformulation is efficient for embedding our problem within a convex SOC programming framework. Due to the objective's incentive to maximize power, \eqref{eq:nu_eta_constraint} is a tight relaxation, i.e., $\nu_{kl}= \sqrt{\eta_{kl}}$ at optimum. Then}, the effective received signal for user $k$ can be denoted by
\begin{equation} \label{GS_termA}
    \mathcal{A}_k(\mathbf{U},  \boldsymbol{\nu}) = \sum\nolimits_{l\in\mathbb{L}} \alpha_l u_{kl}  \nu_{kl} \mathbb{E} [\mathbf{h}_{kl}^T\mathbf{w}_{kl}],
\end{equation} which becomes linear in \( \nu_{kl} \).
Define the additive terms in the denominator of \eqref{eQn:DLSINR_2} as \begin{align} \label{gs_Bk}
        \mathcal{B}_k(\mathbf{U}, \boldsymbol{\nu}) &=  \sum\nolimits_{k'\in \mathbb{K} } \left(      \sum\nolimits_{l\in\mathbb{L}}  u_{k'l} \nu^2_{k'l} |\alpha_l|^2\mathbb{E}\left[|  \mathbf{h}_{kl}^T  \mathbf{w}_{{k'l}}   |^2\right] \right) \\ \nonumber &  + \sum\nolimits_{l \in \mathbb{L}} \beta_l \left (\sum\nolimits_{k' \in \mathbb{K}} u_{k'l} \nu^2_{k'l} \right) \tr \left(\mathbf{R}_{kl}\right)  + {\sigma_z^2}/{p_a},
    \end{align} and the subtracted term as
$\mathcal{C}_k(\mathbf{U}, \boldsymbol{\nu})=\sum\limits_{l \in \mathbb{L}}  u_{kl} \nu^2_{kl}|\alpha_l|^2  \left|\mathbb{E}[ \mathbf{h}_{kl}^T \mathbf{w}_{kl} ] \right|^2$. Thus, the denominator becomes $\mathcal{B}_k(\mathbf{U}, \boldsymbol{\nu}) - \mathcal{C}_k(\mathbf{U}, \boldsymbol{\nu})$.
Introducing a slack variable $\gamma_t$ to represent the common SINR target, the constraint $\gamma_k \geq \gamma_t$ is rewritten to
\begin{align}
    \label{eq:sinr_constraint}
    |\mathcal{A}_k(\mathbf{U},  \boldsymbol{\nu})|^2 \geq \gamma_t \left(\mathcal{B}_k(\mathbf{U}, \boldsymbol{\nu}) -  \mathcal{C}_k(\mathbf{U}, \boldsymbol{\nu}) \right ), \quad \forall k \in \mathbb{K}.
\end{align}
{The constraint in \eqref{eq:sinr_constraint} is still non-convex due to the quadratic term $|\mathcal{A}_k(\mathbf{U}, \boldsymbol{\nu})|^2$. To convexify it, we build a stricter constraint:} 
\begin{equation} \label{eq:sinr_constraint_tight}
   \left( \Re\{\mathcal{A}_k(\mathbf{U},  \boldsymbol{\nu})\} \right)^2 \geq \gamma_t \mathcal{B}_k(\mathbf{U}, \boldsymbol{\nu}), \quad \forall k \in \mathbb{K}.
\end{equation}
\begin{remark}
{Since \( |\mathcal{A}_k(\mathbf{U},  \boldsymbol{\nu})| \geq \Re\{\mathcal{A}_k(\mathbf{U},  \boldsymbol{\nu})\} \) and \( \mathcal{C}_k(\mathbf{U}, \boldsymbol{\nu}) \geq 0 \), it follows that $\left( \Re\{\mathcal{A}_k(\mathbf{U},  \boldsymbol{\nu})\} \right)^2 \geq \gamma_t \mathcal{B}_k(\mathbf{U}, \boldsymbol{\nu}) \Longrightarrow |\mathcal{A}_k(\mathbf{U},  \boldsymbol{\nu})|^2 \geq \gamma_t \left(\mathcal{B}_k(\mathbf{U}, \boldsymbol{\nu}) -  \mathcal{C}_k(\mathbf{U}, \boldsymbol{\nu}) \right )$,
i.e., \eqref{eq:sinr_constraint_tight} is a sufficient (conservative) condition for \eqref{eq:sinr_constraint}. Thus feasibility under \eqref{eq:sinr_constraint_tight} guarantees feasibility under \eqref{eq:sinr_constraint}. }
\end{remark}

\begin{lemma} {
The convexification above guarantees a close approximation because
\[
|\mathcal{A}_k(\mathbf{U},\boldsymbol{\nu})|
\approx
\Re\{\mathcal{A}_k(\mathbf{U},\boldsymbol{\nu})\},
\qquad\text{and}\qquad
\mathcal{C}_k(\mathbf{U},\boldsymbol{\nu}) \ll \mathcal{B}_k(\mathbf{U},\boldsymbol{\nu}).
\] }
\end{lemma}
\begin{IEEEproof} {
The first approximation follows from typical precoder phase alignment (see Sec.~\ref{sec:precoders}): precoder phases are chosen to align with their channels, so $\mathbf{h}_{kl}^T\mathbf{w}_{kl}$ are real — for example, with MR precoding one has $\mathbf{h}_{kl}^T\mathbf{w}_{kl}=\|\mathbf{h}_{kl}\|$. Consequently the sum in \eqref{GS_termA} that defines $\mathcal{A}_k(\mathbf{U},\boldsymbol{\nu})$ is composed mainly of real-valued contributions, making $|\mathcal{A}_k(\mathbf{U},\boldsymbol{\nu})|\approx\Re\{\mathcal{A}_k(\mathbf{U},\boldsymbol{\nu})\}$ a reasonable approximation. 
For the second relation note that
\(
\mathbb{E}\big[|\mathbf{h}_{kl}^T\mathbf{w}_{kl}|^2\big]-\big|\mathbb{E}[\mathbf{h}_{kl}^T\mathbf{w}_{kl}]\big|^2
=\operatorname{Var}(\mathbf{h}_{kl}^T\mathbf{w}_{kl}) \ge 0
\),
so $\mathbb{E}\big[|\mathbf{h}_{kl}^T\mathbf{w}_{kl}|^2\big]\ge\big|\mathbb{E}[\mathbf{h}_{kl}^T\mathbf{w}_{kl}]\big|^2$. In other words, terms collected in $\mathcal{C}_k$ are typically small compared with the squared-mean terms (collected in $\mathcal{B}_k$). This separation becomes more pronounced with extra noncoherent terms $k'\neq k$, especially when $K$ is large. Hence, $\mathcal{C}_k(\mathbf{U},\boldsymbol{\nu})\ll\mathcal{B}_k(\mathbf{U},\boldsymbol{\nu})$ is generally satisfied. Numerical results presented later verify this claim.}
\end{IEEEproof}

Since \( \Re\{\mathcal{A}_k(\mathbf{U},  \boldsymbol{\nu})\} \) is linear in \( \nu_{kl} \), we reformulate \eqref{eq:sinr_constraint_tight} into an SOC form:
\begin{equation}
    \Re\{\mathcal{A}_k(\mathbf{U},  \boldsymbol{\nu})\} \geq \sqrt{\gamma_t} \| \mathbf{b}_k(\mathbf{U}, \boldsymbol{\nu}) \|,  \quad \forall k \in \mathbb{K},
\end{equation}
where \( \mathbf{b}_k(\mathbf{U}, \boldsymbol{\nu}) \) is a vector that stacks the square roots of all \( \mathcal{B}_k(\mathbf{U}, \boldsymbol{\nu}) \)'s terms, namely
\begin{align}
   & \mathbf{b}_k(\mathbf{U}, \boldsymbol{\nu}) = \\  \nonumber
   &\begin{bmatrix} \frac{\sigma_z}{\sqrt{ p_a}}; \;
        \left\{ \nu_{k'l} |\alpha_l| \sqrt{ u_{k'l}  \mathbb{E} [ |\mathbf{h}_{kl}^T \mathbf{w}_{k'l}|^2 ] };\;\nu_{k'l} \sqrt{   u_{k'l} \beta_l  \operatorname{tr} (\mathbf{R}_{kl}) } \right\}_{k' \in \mathbb{K},\; l \in \mathbb{L}} 
    \end{bmatrix}.
\end{align}
{Putting all these elements together, \eqref{eq:maxmin_problem} is transformed {into} a tractable optimization problem:}
\begin{flalign}
    \label{eq:transformed_problem}
    \max_{\mathbf{U},\; \boldsymbol{\nu},\; {\boldsymbol{\eta},}\; \gamma_t} \quad & \gamma_t & \\
    \text{s.t.} \quad & \left\{ \begin{aligned} &\Re\{\mathcal{A}_k(\mathbf{U},  \boldsymbol{\nu})\} \geq \sqrt{\gamma_t} \| \mathbf{b}_k(\mathbf{U}, \boldsymbol{\nu}) \|,\quad \forall k \in \mathbb{K}, & \nonumber \\
    &  \|[\,2\nu_{kl};\;1-\eta_{kl}]\| \;\leq\; 1+\eta_{kl},
     \quad \forall k \in \mathbb{K}, l \in \mathbb{L}, & \nonumber \\
    & \sum\nolimits_{k\in\mathbb{K}} u_{kl}\nu^2_{kl} \leq \frac{1}{ |\alpha_l|^2  + \beta_l  N_a }, \quad \forall l \in \mathbb{L}, \nonumber\\
    &  \nu_{kl} \geq 0,\quad { \eta_{kl} \geq 0, } \quad u_{kl} \in \{0, 1\}, \quad \forall k \in \mathbb{K},\; l \in \mathbb{L}. &\nonumber
    \end{aligned} \right.
\end{flalign}
This problem can be solved by a bisection method on $\gamma_t$, as depicted in Algorithm~\ref{Algorithm_maxminDL}.

\begin{algorithm}
\SetAlgoLined \label{Algorithm_maxminDL}
\DontPrintSemicolon
Initialization: \;
$t \leftarrow 0$,\quad $\gamma^{(0)}_\text{low}\leftarrow 0$, \quad
$\gamma^{(0)}_\text{high} \leftarrow \max_{k\in \mathbb{K}} \left( \frac{p_a\left|  \sum\nolimits_{l\in \mathbb{L}} \alpha_l \mathbb{E} [\mathbf{h}_{kl}^T\mathbf{w}_{kl}] \right|^2}{\sigma_z^2}\right)$\; 
\While{$\gamma^{(t)}_\text{high} - \gamma^{(t)}_\text{low} > \epsilon$}{
    $\gamma_t \leftarrow \frac{1}{2}(\gamma^{(t)}_\text{low} + \gamma^{(t)}_\text{high})$\;   
    \text{Convex Feasibility Check}:  
    \begin{equation*}
    \begin{aligned}
        \text{Find}\quad & \{\mathbf{U},\; \boldsymbol{\nu}\} \\
        \text{s.t.} \quad & \left\{ \begin{aligned} &\Re\{\mathcal{A}_k(\mathbf{U},  \boldsymbol{\nu})\} \geq \sqrt{\gamma_t} \| \mathbf{b}_k(\mathbf{U}, \boldsymbol{\nu}) \|,\quad \forall k \in \mathbb{K}, & \nonumber \\
    &  \|[\,2\nu_{kl};\;1-\eta_{kl}]\| \;\leq\; 1+\eta_{kl},
     \quad \forall k \in \mathbb{K}, l \in \mathbb{L}, & \nonumber \\
    & \sum\nolimits_{k\in\mathbb{K}} u_{kl}\nu^2_{kl} \leq \frac{1}{ |\alpha_l|^2  + \beta_l  N_a }, \quad \forall l \in \mathbb{L}, \nonumber\\
    &  \nu_{kl} \geq 0, \quad u_{kl} \in \{0, 1\}, \quad \forall k \in \mathbb{K},\; l \in \mathbb{L}. &\nonumber
    \end{aligned} \right.
    \end{aligned} 
    \end{equation*}    
    \If{feasible}{
        $\gamma_{\text{low}}^{(t+1)} \leftarrow \gamma_t$,\quad 
        $\gamma_{\text{high}}^{(t+1)} \leftarrow \gamma_{\text{high}}^{(t)}$,\quad
        $\mathbf{U}^* \leftarrow \mathbf{U}$, 
        $\boldsymbol{\nu}^* \leftarrow \boldsymbol{\nu}$\;
    }
    \Else{
        $\gamma_{\text{low}}^{(t+1)} \leftarrow \gamma_{\text{low}}^{(t)}$, \quad
        $\gamma_{\text{high}}^{(t+1)} \leftarrow \gamma_t$\;
    }
    $t \leftarrow t + 1$\;
}
\Return $\{\mathbf{U}^*, \; \boldsymbol{\nu}^*\}$\;
\caption{Joint Optimization Approach}
\end{algorithm}

\section{Computational Cost and Low-Complexity Alternative}

The joint user association and power optimization (JUP) constitutes a mixed-integer SOC program. It involves \( n = L \times K \) binary variables \( u_{kl} \), indicating user-AP associations, and up to \( n \) continuous power variables \( \eta_{kl} \), conditioned on the associated \( u_{kl} = 1 \). In the worst case, all \( 2^n \) combinations of \( \mathbf{U} \) must be explored, resulting in an exponential complexity of \( \mathcal{O}(2^n)\). Such exponential scaling may limit the applicability of this approach to large-scale networks. To address this challenge, we propose a two-stage decoupled strategy:
\begin{enumerate}
    \item \textit{Fixed User Association:} Each user is associated with its geographically closest AP(s), thereby eliminating the binary variables \( u_{kl} \) from the optimization problem.    
    \item \textit{Max–Min Power Control:} With \( \mathbf{U} \) fixed, the problem reduces to optimizing  \( \{ \eta_{kl} \} \) to maximize the minimum achievable SE. Power is allocated only to active associations (i.e., where \( u_{kl} = 1 \)), so the effective number of power variables is smaller than \( L \times K \). 
\end{enumerate}
\begin{table}[t]
\renewcommand{\arraystretch}{1.1}
\caption{Key Simulation Parameters}
\label{tab:sim-params}
\centering
\begin{tabular}{l|l}
\hline \hline
\textbf{Parameter} & \textbf{Value} \\
\hline \hline
Coverage Radius & 1000 m (3GPP Microcell) \\
Path Loss Model & $-30.5 - 36.7\log_{10}(d)$ (dB) \\
Shadow Fading & $\mathcal{N}(0, 4^2)$ dB, Gaussian normal \\
Number of Active Users & $K = 4$ \\
AP Configuration & 32 APs, each with 2 antennas \\
{Power Constraints} & {$p_u = 200$ mW and $p_a = 100$ mW }\\
Noise Spectral Density & $-174$ dBm/Hz \\
Noise Figure & 9 dB \\
System Bandwidth & 5 MHz \\
Antenna Array Type & ULA with half-wavelength spacing \\
Spatial Correlation Model & Gaussian scattering \\
Angular Spread (Std. Dev.) & $30^\circ$ \\
Coherence Interval & $\tau_c = 200$ channel uses \\
Pilot Contamination & $\tau_p = 2$ (2 users per pilot) \\
PA Gain Coefficients & $0.8 + 0.1i \cdot \mathcal{N}(0,1)$, standard normal\\
PA Nonlinearity Factors &  $0.05 + 0.1 \cdot \mathcal{U}(0,1)$, uniform distribution\\
\hline \hline
\end{tabular}
\end{table}
For reference and upper-bound analysis, we compare against the conventional max-min power control \cite{Ref_ngo2017cellfree}, which optimizes all \( L \times K \) power variables. This problem is a standard SOC program with polynomial-time solvability using interior-point methods, corresponding to the complexity on the order of \(\mathcal{O}(n^3)\). By decoupling user association and power control, this low-complexity alternative (named JUP-Lo) enables efficient optimization for larger network deployments.
\begin{table}[ht] 
\caption{Complexity Comparison of Different Optimization Schemes}    \label{tab:complexity_summary}
    \centering
    \begin{tabular}{|c|c|c|c|}
        \hline \hline 
        \textbf{Algorithm} & \textbf{JUP} & \textbf{JUP-Lo} & \textbf{Max-Min Optimization } \\
        \hline
        Complexity & $\mathcal{O}\left(2^{L\times K}\right)$ & $ \mathcal{O}\left({(L\times K)^3}\right)$ & $\mathcal{O}\left({(L\times K)^3}\right)$ \\
        \hline \hline
    \end{tabular}
\end{table}

\begin{figure*}[!tbph]
\centerline{ 
\subfloat[]{
\includegraphics[width=0.3\textwidth]{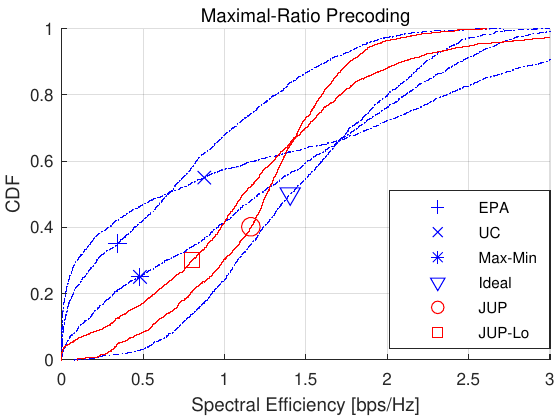}
\label{fig:MR}
}
\hspace{0mm}
\subfloat[]{
\includegraphics[width=0.3\textwidth]{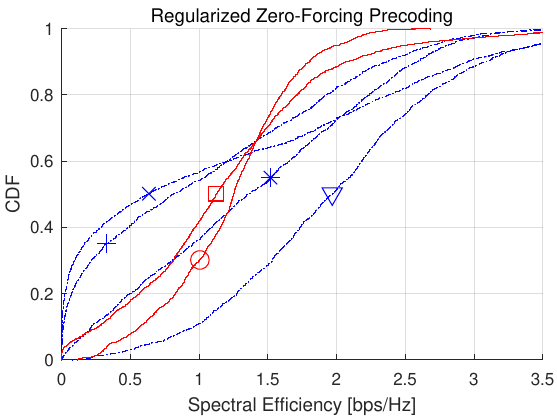}
\label{fig:RZF}
}
\hspace{0mm}
\subfloat[]{
\includegraphics[width=0.3\textwidth]{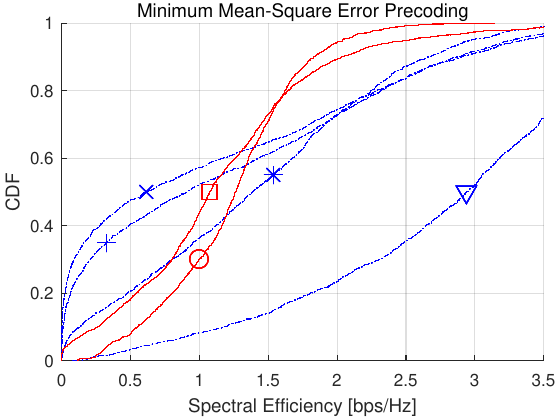}
\label{fig:MMSE}
}
}
\hspace{15mm}
 \caption{The CDF of achievable spectral efficiency under MR, RZF, and MMSE precoding. User fairness (worst-case performance) is represented by the 5th percentile of each curve. All curves use the same marker styles as in Fig. 1a.  }
\label{Fig_performance}
\end{figure*}

\section{Simulation Results and Discussions}

We compare the proposed joint approach with four benchmark schemes: 1) CF configuration with equal power allocation (EPA); 2) CF with max-min power optimization; 3) User-centric (UC) method where each AP serves its nearest user; and 4) Ideal case without PA nonlinearity (i.e., $\alpha_l=1$ and $\beta_l=0$). Key simulation parameters are listed in Table~\ref{tab:sim-params}. The performance under different precoding strategies—MR, RZF, and MMSE—is illustrated in Figs.~\ref{fig:MR}, \ref{fig:RZF}, and \ref{fig:MMSE}, respectively. Each figure shows the cumulative distribution function (CDF) of SE across users. Particular attention is given to the $95\%$-likely SE (i.e., the $5^{th}$ percentile of each CDF), which serves as a key indicator of user fairness and cell-edge performance. The mixed-integer SOC problem in \eqref{eq:transformed_problem} is implemented using CVX \cite{cvx2014} in conjunction with the Gurobi optimizer \cite{gurobi2024}. {Prior to optimization, the expectations in \eqref{eq:transformed_problem} are precomputed by averaging over 100 channel realizations.}

Key observations from the numerical results are as follows:
\begin{itemize}
    \item \textit{PA Nonlinearity Impact:}  The results show that the $95\%$-likely SE under nonlinear PA is very poor, i.e., \text{0.0024}, \text{0.0025}, and \text{0.0020}~bps/Hz for MR, RZF, and MMSE, respectively. In contrast, in the ideal case, the SEs reach \text{0.5635}, \text{0.6542}, and \text{0.6767}~bps/Hz, respectively. This wide performance gap and near-zero SE values underscore the significant adverse impact of PA nonlinearity and highlight the need for PA-resilient design.
    \item \textit{Baseline Limitations:} The applied UC approach yields inferior $95\%$-likely SE (MR: 0.0004, RZF: 0.0003, MMSE: 0.0003 bps/Hz), evident in the CDFs' early plateaus. While reducing fronthaul overhead, its exclusion of non-nearest users creates fairness gaps. The conventional max-min power optimization performs better, achieving \text{0.0289}, \text{0.0956}, and \text{0.0268}~bps/Hz under MR, RZF, and MMSE precoding, respectively. However, using only optimal power optimization remains inadequate—its curves show limited rightward extension compared to ideal cases, indicating its sensitivity to PA nonlinearity.
    \item \textit{Joint Optimization:} { During the simulation, we observe that \(\mathcal{A}_k\) is real for all three precoding schemes and that in most cases \(\mathcal{C}_k/\mathcal{B}_k<10\%\), thereby supporting Lemma~1.}
  Our proposed approach significantly enhances performance (MR: 0.3514, RZF: 0.3602, MMSE: 0.3586 bps/Hz), visualized through \emph{Right-shifted CDF curves} approaching that of ideal PA. It remarkably outperforms the max-min power optimization, achieving gains of approximately \SI{1115}{\percent}, \SI{276}{\percent}, and \SI{1238}{\percent} in $95\%$-likely SE for MR, RZF, and MMSE precoding, respectively. 
\end{itemize}

\begin{figure}[!tbph]
    \centering
    \includegraphics[width=0.32\textwidth]{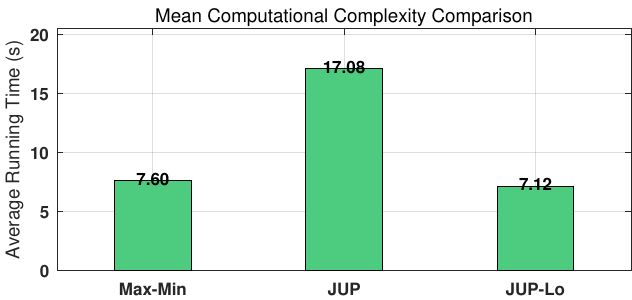}
    \caption{Comparison of computational complexity.}
    \label{fig:complexity}
\end{figure}

{It is not feasible to evaluate complexity numerically as the number of optimization iterations is not fixed}. Therefore, we use the average runtime as a metric to evaluate complexity. On a computer equipped with an Intel i7-4790 processor and 32GB of memory, the average time costs are approximately 17.07, 7.6, and 7.1 seconds for JUP, max-min, and JUP-Lo, respectively, as shown in \figurename \ref{fig:complexity}.  While maintaining lower complexity, the simplified approach still outperforms conventional baselines. It achieves consistent $95\%$-likely performance across all precoding schemes (MR: 0.0531, RZF: 0.0548, MMSE: 0.0567 bps/Hz). These results represent a 2.1$\times$ improvement over max-min power control (0.0268 bps/Hz) and a 178$\times$ enhancement compared to UC  (0.0003 bps/Hz) under MMSE precoding. 

\section{Conclusion}
This letter  {has addressed the challenge of} {investigated} PA nonlinearity in cell-free massive MIMO downlink systems. We proposed a unified analytical framework that models PA-induced distortion across arbitrary linear precoding schemes. Then, a joint optimization approach of user association and power control was {approximated} to {suppress} {alleviate the performance loss raised by} PA nonlinearity. This joint approach, as well as its low-complexity variant, {remarkably} outperforms conventional hardware-agnostic baselines in terms of $95\%$-likely performance, {providing a PA-resilient CF solution.}

\bibliographystyle{IEEEtran}
\bibliography{IEEEabrv,Ref_COML}

\end{document}